\RequirePackage{ifpdf}
\documentclass{JHEP3}

\usepackage{arydshln}

\usepackage{amsfonts,amssymb,amsmath,bm}
\usepackage{slashed}
\usepackage{graphicx}

\newcommand{\Q}{{\cal Q}}

\newcommand{\beq}{\begin{equation}}
\newcommand{\eeq}{\end{equation}}
\newcommand{\beqq}{\begin{equation*}}
\newcommand{\eeqq}{\end{equation*}}
\newcommand\beqa{\begin{eqnarray}}
\newcommand\eeqa{\end{eqnarray}}
\newcommand\beqaa{\begin{eqnarray*}}
\newcommand\eeqaa{\end{eqnarray*}}
\newcommand\bea{\begin{array}}
\newcommand\eea{\end{array}}
\newcommand\beaa{\begin{array}}
\newcommand\eeaa{\end{array}}

\def\XXint#1#2#3{{\setbox0=\hbox{$#1{#2#3}{\int}$ }
\vcenter{\hbox{$#2#3$ }}\kern-.5\wd0}}

\def\hid#1{}

\def\XXint#1#2#3{{\setbox0=\hbox{$#1{#2#3}{\int}$}
\vcenter{\hbox{$#2#3$}}\kern-.5\wd0}}

\newcommand{\nn}{\nonumber}

\newcommand{\neqa}{\nonumber\end{eqnarray}}
\newcommand{\la}[1]{\label{#1}}

\renewcommand{\P}{{\bf P}}

\newcommand{\eq}[1]{(\ref{#1})}

\newcommand{\Tr}{{\rm Tr}}

\newcommand{\hs}{\frac{\sqrt{3}}{2}}
\renewcommand{\d}{\partial}

\newcommand{\<}{{\langle}}
\renewcommand{\>}{{\rangle}}

\newcommand{\re}{\relax{\rm I\kern-.18em R}}

\renewcommand{\sp}{p\hspace{-.40em}/}

\def\su2{{SU(2)}}

\def\[{\left[}
\def\]{\right]}

\def\s{\sigma}

\def\({\left(}
\def\){\right)}
\def\[{\left[}
\def\]{\right]}

\def\<{\langle}
\def\>{\rangle}

\def\cO{{\cal O}}

\def\s*{\ *_{\!\!\!\!\!\!\!\!\!\,_{\,_\text{\scriptsize{sym}}}}}
\def\hs*{\ \hat{*}_{\!\!\!\!\!\!\!\!\!\,_{\,_\text{\scriptsize{sym}}}}}
\def\d{\partial}

\def\i2{\frac{i}{2}}

\def\bQ{{\bf Q}}
\def\bP{{\bf P}}

\def\spi{\relax{\rm \pi\kern-0.5em /}}
\def\sA{\relax{\rm A\kern-0.5em /}}
\def\sp{\relax{\rm p\kern-0.5em /}}
\def\sd{\relax{\rm \d\kern-0.5em /}}
\def\sk{\relax{\rm k\kern-0.5em /}}
\def\sn{\relax{\rm n\kern-0.5em /}}
\def\sl{\relax{\rm l\kern-0.5em /}}
\def\sP{\relax{\rm P\kern-0.7em /}}
\def\sBethe{\relax{\rm \Bethe\kern-0.5em /}}
\def\cN{{\cal N}}

	\renewcommand{\Im}{{\rm Im}}
	\renewcommand{\Re}{{\rm Re}}
	\newcommand{\ofrac}[1]{\frac{1}{#1}}

\def\d{\partial}

    \newcommand{\AD}{{\rm {AdS}}_5\times {\rm S}^5}

\newcommand{\msl}{{\mathfrak{sl}}(2)}
\newcommand{\mpsu}{{\mathfrak{psu}}(2,2|4)}

\title{
Quantum Spectral Curve and the Numerical Solution of the Spectral Problem in $AdS_5/CFT_4$
}

\author{Nikolay Gromov$^{1,2}$,
  Fedor Levkovich-Maslyuk$^{1}$,
  Grigory Sizov$^{1}$ \\
  $^1$King's College London, Department of Mathematics, \\ The Strand, London WC2R 2LS,
  United Kingdom\\ \\
  $^2$ St.Petersburg INP, Gatchina, 188 300, St.Petersburg, Russia
  \qquad\\ \\
  \textit{E-mail:}
  \email{nikgromov$\bullet$gmail.com}, \email{fedor.levkovich$\bullet$gmail.com}, \email{grigory.sizov$\bullet$kcl.ac.uk }\\ \\}

\abstract{
We developed an efficient numerical algorithm for computing the spectrum of anomalous dimensions of
the planar $\cN=4$ Super-Yang--Mills at finite coupling.
The method is based on the Quantum Spectral Curve formalism.
In contrast to Thermodynamic Bethe Ansatz, worked out only for some very special operators, this method is applicable for generic states/operators and is much faster and more precise due to its Q-quadratic convergence rate.

\hspace{0.4cm} To demonstrate the method
 we evaluate the dimensions $\Delta$ of twist operators in $\msl$ sector {{directly}}
 for any value of the spin $S$ including non-integer values.
In particular, we compute the BFKL pomeron intercept in a wide range of the 't Hooft coupling constant with up to $20$ significant figures precision,
confirming two previously known from the perturbation theory orders and giving prediction for several new coefficients.
 Furthermore, we explore numerically a rich branch cut structure for complexified spin $S$.

}

\preprint{}

\begin{document}

\section{Introduction}

Many years of exploring integrable structures in planar ${\mathcal N}=4$ super
Yang-Mills and its $\AD$ string dual have led to a remarkably simple system of equations for the exact spectrum of the theory
 known as Quantum Spectral Curve (QSC) equations.
 They are expected to capture the conformal dimensions/string state energies
at any value of the 't Hooft coupling \cite{PmuPRL,PmuLong}.
 The QSC equations are formulated as a set of Riemann-Hilbert type equations for a few functions.
 As a limiting case this system incorporates the
renowned asymptotic Bethe ansatz, but also
includes all wrapping corrections essential for finite length operators
and perhaps constitutes the ultimate solution of the spectral problem.
The Quantum Spectral Curve has a transparent algebraic
origin related to the underlying $\mpsu$ symmetry of the
problem. It was derived in \cite{PmuPRL,PmuLong} from the discrete integrability of
Hirota dynamics underlying the Y-system enhanced with a specific analyticity condition worked out in
\cite{Gromov:2009tv,Bombardelli:2009ns,Gromov:2009bc,Arutyunov:2009ur,Cavaglia:2010nm}.
In contrast to some explicitly
known integral forms of the Y-system \cite{Gromov:2009bc,Gromov:2011cx,Balog:2012zt,Suzuki:2011dj,Bajnok:2013wsa}
the QSC is applicable for any local operator or string state.

The striking simplicity of the QSC formulation has already led to a rapidly growing body of exact as well as perturbative results.
First, in \cite{PmuPRL,Gromov:2014bva} it was pointed out that it can be solved analytically in a near-BPS regime,
e.g. for expansion in the coupling $\lambda$ or in the spin $S$, in particular in \cite{Gromov:2014bva} the
anomalous dimensions for twist operators in the $\msl$ sector were computed at any coupling to order $S^2$,
giving new analytical predictions at strong coupling for some local operators and the BFKL pomeron intercept.
As an extension of the observation of \cite{PmuPRL}
powerful techniques for expansion in the coupling were developed in \cite{Marboe:2014gma,Marboe:2014sya}
which led, impressively, to 10-loop anomalous dimensions for numerous operators,
as well as 6-loop predictions at any $S$ for twist-2 operators.
Finally, in \cite{Alfimov:2014bwa} the analytic continuation to the BFKL region $S\sim-1$ was explored,
and the leading order BFKL equation
in SYM was derived in this approach.
It is also important to mention that the QSC method was also applied
to ABJM theory \cite{Cavaglia:2014exa} opening a way for various explicit calculations in particular to that of the
interpolation function $h(\lambda)$ \cite{Gromov:2014eha}, the mysterious extra component of the spectral problem in this model.

Even though many explicit analytic results are now available both at strong and weak coupling, one important range of applications of the QSC that has remained unexplored till now is the
numerical investigation of the spectrum at finite coupling.

Previous numerical methods based on TBA, even limited to a few operators\footnote{
only for a few operators the complicated structure of the ``driving terms" was deduced explicitly in a closed form.
Even for those operators the driving terms may change depending on the value of the coupling.
}, low precision and slow convergence rate
gave, nevertheless, several highly important results, allowing, in particular,
the computation of the anomalous dimension of a nonprotected (Konishi) operator in
a planar 4d theory at finite coupling \cite{Gromov:2009zb}. Numerics also gave a prediction for the strong coupling Konishi anomalous dimension which was later confirmed by several methods \cite{Roiban:2011fe,Vallilo:2011fj,Gromov:2011de,Basso:2011rs,Gromov:2014bva,Gromov:2011bz,Frolov:2010wt,Frolov:2013lva} .
The main goal of the present work is to remove the limitations of the previously known methods by
developing an algorithm for a numerical solution of the QSC.

The low precision and performance of the TBA-like
approach was mainly due to the complicated infinite system of equations and cumbersome integration kernels.
The QSC includes only a few unknown functions and thus can be expected to give highly
precise numerical results. However, the QSC equations are functional equations
supplemented with some
analyticity constraints of a novel type which makes it a priori not a trivial task to develop a robust numerical approach.

In this paper we propose an efficient method to solve the QSC numerically and illustrate our method by a few examples. Among the several equivalent formulations of the QSC we identified the equations
which are best-suited for numerical solution\footnote{one may call this sub-system of equations as $\bP\omega$-system, in contrast to previously used $\bP\mu$-system or $\bQ\omega$-system}. We implemented our algorithm in
{\it Mathematica} and were able to get a massive increase in efficiency compared to the TBA or FiNLIE systems \cite{Gromov:2009zb,Gromov:2011cx,Frolov:2010wt,Balog:2012zt}.
With one iteration taking about $2$ seconds we only need $2-3$ iterations (depending on the starting points) to reach at least $10$ digits of precision.
Quite expectedly, the precision gets lost for very large values of the 't Hooft coupling.
Nevertheless, without any extra effort we reached $\lambda\sim 1000$ keeping a good precision, which should be more than enough for any practical goal.

Not only does our approach work for any finite length single trace operator and in particular for any value of the spin,
it also works with minimal changes even away from integer quantum numbers! We demonstrate this in the particularly interesting case of the $\msl$ twist-2 operators. Their anomalous dimension analytically continued to complex values of the spin $S$ is known to have a very rich structure, in particular the region $S\simeq-1$ is described by BFKL physics.
As we show, within the framework of QSC it is not hard to specify any value of the Lorentz spin $S$
as the conserved charges enter the equations through the asymptotics which can in principle take any complex values.
Then we can compute the analytically continued scaling dimension $\Delta$ {\textit {directly}} for complex $S$ (or even interchange
their roles and study $S$ as a function of $\Delta$).
The result of this calculation can be seen on Fig.~\ref{fig:sdelta}.

\FIGURE[ht]
{
\label{fig:sdelta}

    \begin{tabular}{cc}
    \includegraphics[natwidth=960,natheight=720,scale=0.7]{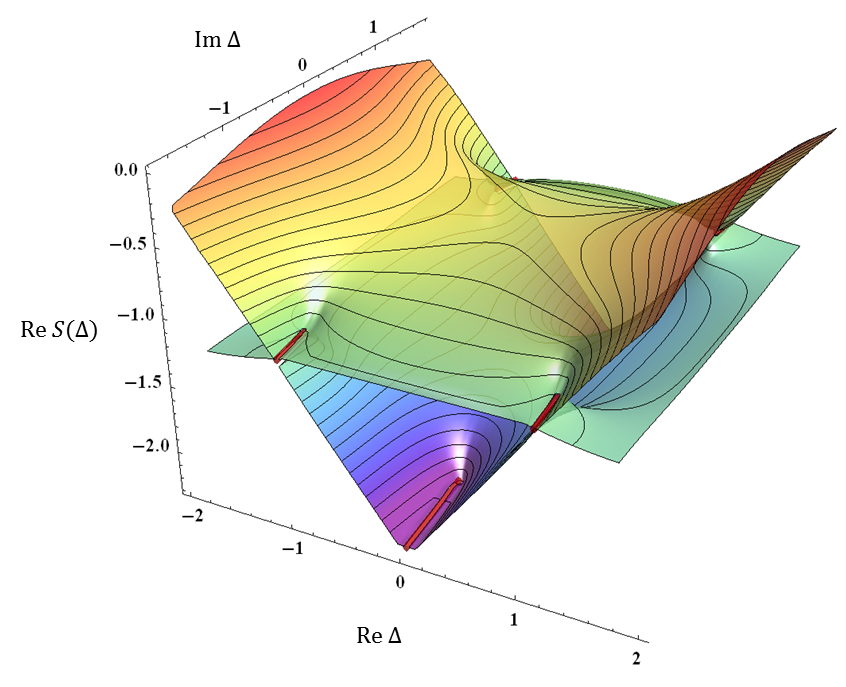}\\
    \end{tabular}
\caption{\textbf{Riemann surface of the function $S(\Delta)$ for twist-2 operators.} Plot of the real part of $S(\Delta)$ for complex values of $\Delta$, generated from about 2200 numerical data points for $\lambda\approx 6.3$. We have mapped two Riemann sheets of this function. The thick red lines show the position of cuts. The upper sheet corresponds to physical values of the spin. Going through a cut we arrive at another sheet containing yet more cuts.}
}

Let us stress that the algorithm is very simple and mainly consists of elementary matrix operations. As such it can be easily implemented on
various platforms. In particular, we believe the performance could be increased by a few orders
with a lower level, e.g. C++, implementation. In this paper we mostly aim to demonstrate our algorithm, prototyped in {\it Mathematica}. To illustrate how the algorithm works, as an attachement to this paper we provide a Mathematica notebook with a simple implementation of our method.



Finally, to improve the performance of our method we used the further simplification of the QSC obtained in \cite{Gromov:2015vua}, which allows us to eliminate auxiliary functions $\omega_{ij}$ from our algorithm and close the equations using Q-functions only (we demonstrate this for the $sl(2)$ sector states).

The structure of the paper is as follows. In section 2 we introduce the Quantum Spectral Curve and discuss the key equations for the numerical implementation. In section 3 we describe how our algorithm works
in general. We also demonstrate our method on specific examples for twist-2 operators in the $\msl$ sector,
starting with physical operators (such as the well studied Konishi operator).
Then in section 4 we explain how to extend the method to non-integer values of the spin. We discuss Fig.~\ref{fig:sdelta} in a bit more detail,
and also describe our high precision evaluation of the BFKL pomeron intercept\footnote{i.e. the value of $S$ for which $S(\Delta)=0$}.
The final section contains our conclusions and speculations.

\section{Review of the AdS/CFT Quantum Spectral Curve}

Let us start by introducing the main equations of the Quantum Spectral Curve framework. For a full review of the QSC we refer the reader to \cite{PmuLong},
but we will try to make the paper self-contained.

For integrable spin chains Baxter Q-polynomials play a central role. They carry complete information about the state and the spectrum
and can also be thought of as wave functions of the model.
They are completely determined by a set of functional relations (known as Q-system) and by the polynomiality condition.

In ${\cal N}=4$ SYM the situation is similar.
Its spectrum of anomalous dimensions is described
by a set of Q-functions satisfying the same Q-system relations.
However, in this model the Q-functions are no longer polynomials, but are analytic functions of the spectral parameter $u$ with branch cuts.
The positions of the branch points depend on the `t Hooft coupling $\lambda$: they are situated at $\pm 2g+i n,\;n\in{\mathbb Z}$, where $g=\frac{\sqrt{\lambda}}{4\pi}$.
In the limit when $\lambda$ is small the system reduces to the usual $\mathfrak{psu}(2,2|4)$ Heisenberg spin chain.
For the class of functions with cuts the Q-system alone is not constraining enough and one has
to specify the monodromies of the Q-functions in order to close the system.
These monodromies were found in \cite{PmuPRL} and
take the form of a Riemann-Hilbert problem as described below. The resulting set of equations found in \cite{PmuPRL,PmuLong} is known under the name Quantum Spectral Curve (QSC).

\FIGURE[ht]{
\includegraphics[scale=0.18]{pq2v2}
\caption{ $\bP_a$ and $\bQ_i$ have one cut on the real axis in the representations with short and long cuts respectively. The ellipse shows the region of convergence of the series \label{fig:PQ}\protect\eq{param}.}
}

The Q-system of ${\cal N}=4$ SYM is composed of $2^8$ Q-functions. However, the algebraic relations between them allow us to choose a much smaller subset, which will be complete in the sense that the rest of Q-functions can be generated from the selected ones algebraically. A convenient choice for such a subset consists of $4+4$ functions $\bP_a(u)$ and $\bQ_i(u)$
($a,i=1,\dots,4$). One can say that $\bP_a$ describe the $S^5$ degrees of freedom whereas $\bQ_i$
correspond to the $AdS_5$ part.
A particularly nice property of $\bP$'s is that they have only two branch points
at $\pm 2g$ when they are connected by a ``short'' cut $[-2g;2g]$ (see Fig. \ref{fig:PQ}).
This means that there are more branch points on the next sheet, but for this choice of the cut they do not appear on the first sheet.
Very similarly $\bQ$'s have only two branch point on the main sheet if the cut is taken to go through infinity. In a sense this reflects the non-compactness of the
$AdS_5$ part of the space.

Whereas the coupling determines the position of the branch points, the quantum numbers of the state are specified through  the large
$u$ asymptotics of Q-functions. $\bP_a$ encode the compact bosonic subgroup $SO(6)$ quantum numbers $(J_1,J_2,J_3)$,
while $\bQ_i$ give the $SO(4,2)$ charges $(\Delta,S_1,S_2)$, which include the conformal dimension of the state $\Delta$.
Explicitly
\beq
	\bP_a\sim A_au^{-\tilde M_a}, \ \ \bQ_i\sim B_iu^{\hat M_i-1},
\label{asymptPQ}
\eeq
where
\begin{align}
\label{relMta}
\tilde M_a=
&\left\{\frac{J_1+J_2-J_3+2}{2}
  ,\frac{J_1-J_2+J_3}{2}
   ,\frac{-J_1+J_2+J_3+2}{2}
   ,\frac{-J_1-J_2-J_3}{2}
   \right\}\ ,\\
\hat M_i=&
\left\{
\frac{\Delta -S_1-S_2+2}{2} ,
\frac{\Delta +S_1+S_2}{2}
   ,
\frac{-\Delta-S_1+S_2+2}{2} ,
\frac{-\Delta+S_1-S_2}{2} \right\}\ .
\label{M-ass}\end{align}

Note that for the numerical implementation $\bP_a$ are more handy, in the sense that
they can be expressed as a series in the Zhukovsky variable $x(u)$ defined by $u=g(x+1/x)$,
\beq\label{param}
\bP_a(u)=\sum_{n=\tilde M_a}^\infty \frac{c_{a,n}}{x^n(u)}\;.
\eeq
This series is convergent everywhere on the upper sheet and also in an elliptic region around the cut on the next sheet (see Fig.~\ref{fig:PQ}).
A similar parametrization for $\bQ_i$ will not cover even the upper sheet.
Fortunately, in the whole set of $2^8$ Q-functions there are other $4$ functions with
one single cut, which are denoted as $\bP^a(u)$, $a=1,\dots,4$.
Together with $\bP_a(u)$ they also form a complete set of Q-functions. In particular, one can reconstruct $\bQ_i$ from them.
The procedure for this, which will be crucially important in our numerical implementation, is the following:
\begin{itemize}
\item{
Find a set of $16$ functions ${\cal Q}_{a|i}$, satisfying
\beq\la{Qai}
{\cal Q}_{a|i}(u+\tfrac i2)-{\cal Q}_{a|i}(u-\tfrac i2)=-
\bP_a(u)
\bP^b(u)
{\cal Q}_{b|i}(u+\tfrac i2)\ \ .
\eeq
Note that this is a $4$-th order finite difference equation, which entangles all
${\cal Q}_{a|i}$ with fixed $i$. Different values of $i$ label the $4$ linearly independent solutions of this equation. One could also equivalently use ${\cal Q}_{b|i}(u-\tfrac i2)$ in place of ${\cal Q}_{b|i}(u+\tfrac i2)$ in the r.h.s., due to the constraint \cite{PmuLong}
\beq
\label{PPzer}
	\bP_a\bP^a=0\ .
\eeq
This constraint also fixes some of the coefficients $c_{a,n}$.
}
\item{
The matrix ${\cal Q}_{a|i}$ can then be used to pass to $\bQ_i$ from $\bP^a$'s,
\beq\la{QfromQai}
\bQ_i(u)=-\bP^a(u)\;{\cal Q}_{a|i}(u+i/2)\ .
\eeq
}
The equations \eq{Qai} and \eq{QfromQai} are simply two of the Q-system relations as explained in \cite{PmuLong}.
We also introduce a matrix ${\cal Q}^{a|i}$ such that ${\cal Q}^{a|i}{\cal Q}_{a|j}=-\delta^{i}_{j}$ and use it to define $\bQ$'s with an upper index:
\beq
\la{QupfromQai}
\bQ^i(u)=+\bP_a(u)\;{\cal Q}^{a|i}(u+i/2)\;.
\eeq

Note that since ${\cal Q}_{a|i}(u)$ is analytic in the upper-half-plane we can also analytically continue these
relations around the branch point at $u=2g$ to get
\beqa\la{QfromQait}
\tilde\bQ_i(u)=-\tilde\bP^a(u)\;{\cal Q}_{a|i}(u+i/2)\\
\tilde\bQ^i(u)=+\tilde\bP_a(u)\;{\cal Q}^{a|i}(u+i/2)
\eeqa
where the tilde denotes analytic continuation to the next sheet.

Let us also mention that large $u$ asymptotics of $\bP^a$ and $\bQ_i$ read \cite{PmuLong}
\beq
	\label{asymptPQ2}
		\ \bP^a\sim A^au^{\tilde M_a-1},\
	\bQ^i\sim B^iu^{-\hat M_i}\ .
\eeq

\end{itemize}

Since $\bQ_i$ can now be recovered from $\bP_a$ and $\bP^a$ it is not surprising that actually all  information we need,
in particular all the charges (including those in $AdS_5$), are encoded in $\bP$'s alone, through
\beq
	\bP_a\sim A_au^{-\tilde M_a}, \ \bP^a\sim A^au^{\tilde M_a-1}, \ \
	A^{a_0}A_{a_0}=i \frac{\prod_j(\tilde M_{a_0}-\hat M_{j})}{\prod_{b\neq a_0}(\tilde M_{a_0}-\tilde M_{b})}\la{largeA},
\eeq
where $\hat M_j$ and $\tilde M_a$ are defined in \eq{relMta}, \eq{M-ass} and $a_0$ takes values $1,2,3,4$ but there is no summation over $a_0$ in l.h.s.
In particular, one can extract $\Delta$ from the last equation.

The coefficients $c_{a,n}$ and corresponding coefficients $c^{a,n}$ of the expansion of $\P^a(u)$ need to be found. The constraint \eq{PPzer} fixes some of them (for example, we can use it to fix all $c_{1,n}$). The condition \eq{largeA} gives the leading coefficients $c_{a,\tilde M_a}$. The remaining coefficients should be fixed from the analyticity constraints on $\bP$'s as prescribed by QSC.
Let us describe these constraints.
The analytic continuation of $\bP_a$ to the second sheet, which we denote by $\tilde\bP_a$,
in terms of our ansatz \eq{param} becomes simply
\beq\la{paramt}
\tilde\bP_a(u)=\sum_{n=\tilde M_a}^\infty {c_{a,n}}{x^n(u)}\;.
\eeq
According to \cite{PmuLong} we should have
\beq
\tilde \bP_a(u)=\mu_{ab}(u)\bP^b\;\;,\;\;\tilde \bP^a(u)=\mu^{ab}(u)\bP_b(u)\la{Pmu}
\eeq
where $\mu_{ab}(u)$ is an antisymmetric matrix with unit Pfaffian, $i$-periodic as a function with long cuts, with the discontinuity fixed in terms of $\bP_a$
\beq
\tilde\mu_{ab}(u)-\mu_{ab}(u)=\bP_a\tilde \bP_b-\tilde\bP_a\bP_b\;. \la{muab}
\eeq

\FIGURE[ht]{
\includegraphics[scale=0.18]{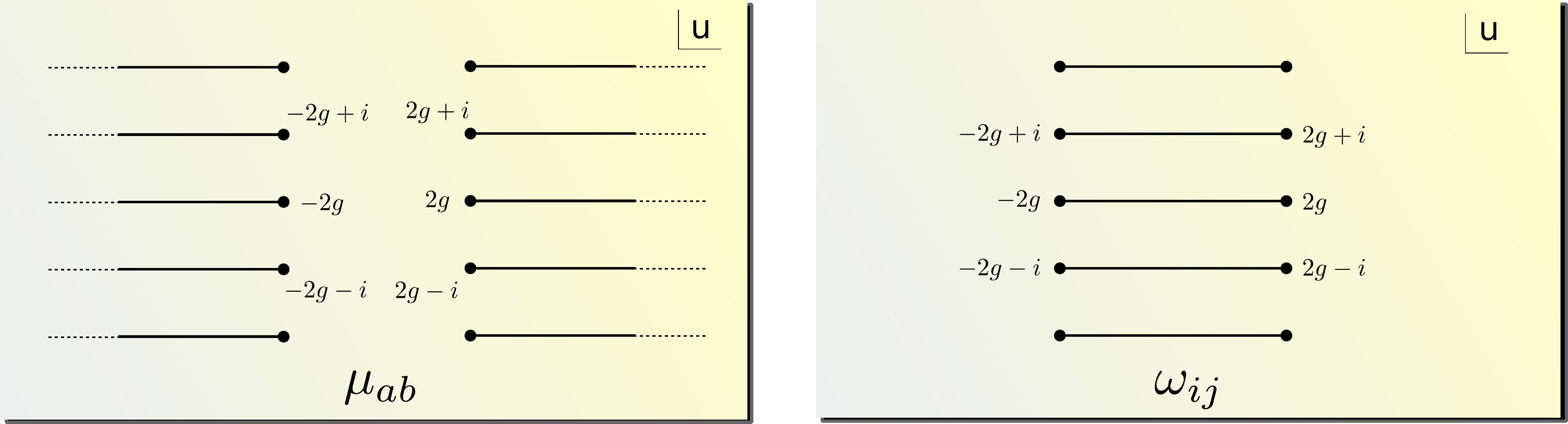}
\caption{$\mu_{ab}$ is periodic as a function with long cuts, and $\omega_{ij}$ as a function with short cuts.}
}

Knowing the r.h.s. of \eq{muab} it is straightforward to reconstruct $\mu_{ab}$ itself using the spectral representation of $\mu_{ab}$
\beq
\mu_{ab}(u)=\frac{i}{2}\[\;\int\limits_{2g}^\infty dv+\int\limits_{-\infty}^{-2g}dv\] \coth(\pi (u-v))\[\tilde \bP_a(v)\bP_b(v)-\bP_a(v)\tilde\bP_b(v)\]+{\rm periodic},
\label{muabintegral}
\eeq
In theory one could reconstruct $\mu_{ab}$ for some fixed coefficients $c_{a,n}$ from \eq{muab} and then impose \eq{Pmu} to fix the unknown $c_{a,n}$.
However, in practice that is very hard to do as the power series \eq{param} is only convergent inside the region shown on Fig \ref{fig:PQ}. The problem is that this region does not cover
the entire cut of $\mu$, which stretches to infinity. In other words, it would be very hard to reconstruct $\mu_{ab}$ from some given coefficients
$c_{a,n}$ in this direct way.

We found that it is much more advantageous to close the analyticity conditions at the level of $\bQ_i$, which obey very similar equations. The rule is quite simple -- one has to interchange short and long cuts.
That is, we have to introduce an $i$-periodic with {\it short} cuts function $\omega_{ij}(u)$ such that
\beq
\label{muab1}
\tilde \bQ_i=\omega_{ij}\bQ^j\;\;,\;\;\tilde \bQ^i=\omega^{ij}\bQ_j
\;\;,\;\;
\eeq
\beq
\tilde\omega_{ij}-\omega_{ij}= \bQ_i\tilde\bQ_j-\tilde\bQ_i\bQ_j\;. \la{omegad}
\eeq
with ${\rm Pf}\;\omega=1$.
Now we see that to recover $\omega_{ij}$ one only needs to know its discontinuity on the interval $[-2g,2g]$,
which is completely inside the region of convergence on Fig.~\ref{fig:PQ}! Thus the discontinuity of $\omega$
can be expressed in terms of the coefficients $c_{a,n}$
via \eq{QfromQai} and \eq{QfromQait}, provided we know how to solve \eq{Qai} for arbitrary $\bP_a$ and $\bP^a$.
In the next section we will describe an algorithm which allows to solve \eq{Qai} very efficiently and then find the coefficients $c_{a,n}$, which yields the solution of the QSC. We will also show that actually finding $\omega_{ij}$ is not necessary and we can close the system in terms of just $\bP_a,\bQ_i$ and ${\cal Q}_{a|i}$, thus speeding up the calculations.

\section{Description of the Method}
\label{sec:Solving}
\subsection{Step 1: Solving the equation for ${\cal Q}_{a|i}$}
As we explained above the quantity ${\cal Q}_{a|i}$ is at the heart of our procedure. In this section we will demonstrate how this
set of $16$ functions can be found for arbitrary $\bP_a$ and $\bP^a$. In this procedure the precise ansatz for $\bP$ is not important. However, as we will see later,
we should be able to compute $\bP_a(u)\bP^b(u)$ on the upper sheet for $u$ with large imaginary part. Of course, having the ansatz
in the form of a (truncated) series expansion \eq{param} we can easily evaluate it everywhere on the upper sheet
numerically very fast.

The process of finding ${\cal Q}_{a|i}$ is divided into two parts.
Firstly, we find a good approximation for ${\cal Q}_{a|i}$ at some $u$ with large imaginary part (in the examples we will need ${\rm Im}\; u\sim 10-100$).
At the next step we apply to this large $u$ approximation of ${\cal Q}_{a|i}$  a recursive procedure which produces ${\cal Q}_{a|i}$ at $u\sim 1$.

\paragraph{Large $u$ approximation.} For ${\rm Im}\; u\sim 10-100$ we can build the solution of \eq{Qai} as a  $1/u$ expansion.
This is done by plugging the (asymptotic) series expansion of ${\cal Q}_{a|i}$ into \eq{Qai}:
\beq
\label{Qlarge}
{\cal Q}_{a|i}(u)=u^{\hat M_i-\tilde M_a}\sum_{n=0}^{N}\frac{B_{a|i,n}}{u^n}.
\eeq
where $N$ is some cutoff (usually $\sim 10-20$).
This produces a simple linear problem for the coefficients $B_{a|i,n}$, which can be even solved analytically to a rather high order.
The leading order coefficients of ${\cal Q}_{a|i}$ can be chosen arbitrarily. After that the linear system of equations becomes non-homogenous and
gives a unique solution in a generic case.\footnote{The matrix of this system may become non-invertible
unless some constraint (which is not hard to find) on the coefficients $c_{a,n}$ is satisfied. This constraint is fulfilled trivially for the even in the rapidity plane operators considered in the next section.
There is also no such problem for the situation with generic twists (similar to $\beta-$ or $\gamma$-deformations, see the review \cite{Zoubos:2010kh}). Adding twists should correspond \cite{PmuLong} to allowing exponential factors $e^{\alpha_a u}, e^{\beta_i u}$ in the asymptotics of $\bP_a$ and $\bQ_i$, making everything less degenerate and providing a useful regularization.

}

\paragraph{Finite $u$ approximation.}
Once we have a good approximation at large $u$ we can simply use the equation \eq{Qai}
to recursively decrease $u$. Indeed defining a $4\times 4$ matrix
\beq
{U_{a}}^b(u)=\delta_a^b+\bP_a(u)\bP^b(u)
\eeq
we have
\beq
{\cal Q}_{a|i}(u-\tfrac i2)={{U}_{a}}^b(u){\cal Q}_{b|i}(u+\tfrac i2).
\eeq
Iterating this equation we get, in matrix notation
\beq
{\cal Q}_{a|i}(u-\tfrac i2)=\[U(u)U(u+i)\dots U(u+i N)\]_{a}{}^b\;{\cal Q}_{b|i}(u+i N+\tfrac {i}2)\;.
\eeq
For large enough  $N$ we can use the large $u$ approximation \eq{Qlarge} for
${\cal Q}_{b|i}$ in the r.h.s. As a result we obtain the functions ${\cal Q}_{a|i}$ for finite $u$ with high precision.

\subsection{Step 2: Recovering $\omega_{ij}$}
\la{sec:Recovering}
Now when we have a good numerical approximation for ${\cal Q}_{a|i}(u)$ we can compute
$\bQ_i$ and $\tilde \bQ_i$ which through the discontinuity relation \eq{omegad} will yield us $\omega_{ij}$. 
 
Let us also note that, as it was argued in our paper \cite{Gromov:2015vua}, one can in fact close the QSC equations without calculating $\omega_{ij}$ (this was shown explicitly for the symmetric $sl(2)$ sector states). This makes it possible to further speed up our numerical procedure as we will describe in detail in section \ref{sec:Application}. In the current section for completeness we will present the 
procedure without this shortcut, as for some applications it could turn out to be useful as well.

%

 
One can recover $\omega_{ij}$ from its discontinuity \eq{omegad} modulo an analytic function, as due to $i$-periodicity of $\omega_{ij}$ one can write its spectral representation as
\beq\la{int}
\omega_{ij}(u)=\frac{i}{2}\int\limits_{-2g}^{2g}dv \coth(\pi (u-v))\[\tilde \bQ_i(v)\bQ_j(v)-\bQ_i(v)\tilde\bQ_j(v)\]+\omega^0_{ij}(u)
\eeq
where the ``zero mode" $\omega^0_{ij}(u)$ is the analytic part of $\omega_{ij}$ ~--- it has to be periodic, antisymmetric in $i,j$ and
should not have cuts. We will fix it below. One can check directly that the term with the integral in the above expression is periodic and moreover due to the pole of the $\coth$ function it has precisely the correct discontinuity on the real axis to satisfy \eq{omegad}. Therefore this term gives a particular $i$-periodic solution of \eq{omegad}, while the $\omega^0_{ij}(u)$ is the general solution of the same equation \eq{omegad} with r.h.s. set to zero, so that \eq{int} indeed gives the general solution of \eq{omegad}. 

We note that we only need to know values of $\bQ$ and $\tilde\bQ$ on the cut. In our implementation we use a finite number of sampling points on the
cut given by zeros of Chebyshev polynomials. One can then fit the values of $\tilde\bQ_i\bQ_j-\bQ_i\tilde\bQ_j$ at those points with a polynomial times the square root $\sqrt{u^2-4g^2}$.
After that we can use precomputed integrals of the form
$\int_{-2g}^{2g} \coth(\pi (u_i-v)) v^n \sqrt{v^2-4g^2}dv$ to evaluate \eq{int} with high precision by a simple matrix multiplication,
which produces the result at the sampling points $u_A$ in an instant.

One more point to mention here is that in our implementation we only compute $\omega_{ij}^{reg}=\frac{1}{2}(\omega_{ij}-\tilde \omega_{ij})$
at the sampling points to avoid the problem of dealing with the singularity of the integration kernel.
Note that
$\omega^{reg}_{ij}$ can be used instead of $\omega_{ij}$ in the equations like \eq{muab1}, because the difference is proportional to $\bQ_i\bQ^i$
which is zero similarly to \eq{PPzer}, as can be shown by combining \eq{PPzer} with \eq{QfromQai}, \eq{QupfromQai}.

\paragraph{Finding zero modes.}
\la{sec:Findingzm}
It only remains to fix $\omega^0_{ij}(u)$. First we notice that for all physical operators $\omega_{ij}$ should not grow faster than constant at infinity
\cite{PmuLong}. As the integral part of \eq{int} does not grow either and since $\omega^0_{ij}(u)$ is $i$-periodic it can only be a constant.
To fix this constant we use the following observation \cite{PmuLong}: the constant matrix $\alpha_{ij}^+$ which $\omega_{ij}$
approaches at $u\to+\infty$ and the constant matrix $\alpha_{ij}^-$ which it reaches at $u\to-\infty$ are restricted by the quantum numbers \cite{PmuLong}.
To see this we can pick some point on the real axis far away from the origin and shift it slightly up into the complex plane,
then from \eq{muab1} we have
\beq\la{omr}
\omega_{ij}\bQ^j(u+i0)=\alpha^+_{ij}\bQ^j(u+i0)=\tilde\bQ_i(u+i0)=\bQ_i(u-i0).
\eeq
Similarly for $-u$ we get
\beq\la{oml}
\alpha^-_{ij}\bQ^j(-u+i0)=\bQ_i(-u-i0).
\eeq
Next, notice that since $\bQ^j$ is analytic everywhere except the cut on the real axis, it can be replaced by its asymptotics
above the real axis, i.e. $\bQ^j(u+i0)\sim B^j u^{-\hat M_j}$,
and also $\bQ^j(-u+i0)\sim B^j u^{-\hat M_j}e^{-i\pi \hat M_j}$, as we find from the previous expression by a rotation by $\pi$ in the complex plane.
As a result we get the asymptotics of $\bQ_i$ at infinities and slightly below the real axis
\beqa\la{seco}
\bQ_i(u-i0)=\alpha^+_{ij}B^j u^{-\hat M_j}\;\;,\;\;\bQ_i(-u-i0)=\alpha^-_{ij}B^j u^{-\hat M_j}e^{-i \pi \hat M_j}\;.
\eeqa
Using that they are related by the analytic continuation in the lower half plane the first equation also gives
\beq
\bQ_i(-u-i0)=\alpha^+_{ij}B^j u^{-\hat M_j}e^{+i \pi \hat M_j}\ .
\eeq
Combining this with \eq{seco} we get a relation between the constant asymptotics of $\omega$ at the two infinities
\beq
\alpha^+_{ij}=\alpha^-_{ij}e^{-2i\pi \hat M_j}\;.
\eeq
At the same time from \eq{int} we get
\beq \la{alphaI}
\alpha^\pm_{ij}=\pm I_{ij}+\omega^0_{ij}\;\;,\;\;I_{ij}\equiv \frac{i}{2}\int\limits_{-2g}^{2g} dv\[ \tilde\bQ_i(v)\bQ_j(v)-\bQ_i(v)\tilde\bQ_j(v)\],
\eeq
which implies that
\beq
\omega^0_{kl}=-i I_{kl}\cot \pi \hat M_l.
\la{omegaI}
\eeq
We see that the zero modes can be also computed from the values of $\bQ$ and $\tilde\bQ$
on the cut.

Note also that the r.h.s. is not explicitly antisymmetric. Imposing the antisymmetry gives
\beq\la{antisymm}
 I_{kl}(\cot \pi \hat M_l-\cot \pi \hat M_k)=0,
 \eeq
 so either $I_{kl}=0$
or $\cot \pi \hat M_l=\cot \pi \hat M_k$.
As $\textrm{Pf}\;\omega=1$, all $I_{kl}$ can not be equal to zero simultaneously. Having $I_{kl}$ non-zero implies quantization of charges: for example, the choice $I_{12}\neq 0$ and $I_{34}\neq 0$, which is
consistent with perturbative data, requires $\cot \pi \hat M_1=\cot \pi \hat M_2$ and $\cot \pi \hat M_3=\cot \pi \hat M_4$, and so $S_1,S_2$ have to be integer or half integer.
In section \ref{sec:nonint} we will see how to relax this condition and do an analytic continuation in the spin $S_1$ to the whole complex plane.

\subsection{Step 3: Reducing to an optimization problem}
\la{sec:Finding}
Having $\omega_{ij}$ and $\bQ_{a|i}$ at hand we can try to impose the remaining equations of the QSC
\eq{muab1}. We notice that there are two different ways of computing $\tilde \bQ_i$, which should give the same result
when we have a true solution: \eq{QfromQait} and \eq{muab1}. Their difference, which at the end should be zero, is
\beq\label{errorfunction}
F_i(u)=\tilde\bP^a(u)\;{\cal Q}_{a|i}(u+i/2)+\omega_{ij}(u)\;{\cal Q}^{a|j}(u+i/2)\bP_a(u)\ .
\eeq
The problem is now to find $c_{a,n}$ for which $F_i(u)$ is as close as possible to zero.
Here we have some freedom in how to measure its deviation from zero, but in our implementation we use the sum of squares of
$F_i$ at the sampling points $u_A$.
Then the problem reduces to the classical optimization problem
of the least squares type. In our implementation we found it to be particular efficient to use the
Levenberg-Marquardt algorithm (LMA), which we briefly describe in the next section.
The LMA is known to have a Q-quadratic convergence rate, which means that the error $\epsilon_n$
decreases with the iteration number $n$ as fast as $e^{-c\; 2^n}$.
The convergence is indeed so fast that normally it is enough to do $2$ or $3$ iterations to get the result with $10$ digits precision.
We give some examples in the next section.

\paragraph{Levenberg-Marquardt algorithm}
Our problem can be reformulated as follows: given a vector function  $f_i(c_a)$ of a set of variables $c_a$
(which we can always assume to be real)
find the configuration which minimizes
\beq
\label{Sdef}
	S(c_a)\equiv \sum\limits_i |f_i(c_a)|^2\ .
\eeq
Assuming we are close to the minimum we can approximate $f_i$ by a linear function:
\beq
f_i(\tilde c_a)\simeq f_i(c_a)+(\tilde c_a-c_a)J_{ai}\;\;,\;\;J_{ai}\equiv \d_{c_a}f_i(c_a)
\eeq
which gives the following approximation for $S(\tilde c_a)$:
\beq
S(\tilde c_a)=\[f_i(c_a)+(\tilde c_a-c_a)J_{ai}\]
\[\bar f_i(c_a)+(\tilde c_a-c_a)\bar J_{ai}\]
\eeq
The approximate position of the minimum is then at $\d_{\tilde c_a} S=0$ for which we get
\beq
J_{ai}\[\bar f_i(c_a)+(\tilde c_a-c_a)\bar J_{ai}\]
+\[f_i(c_a)+(\tilde c_a-c_a)J_{ai}\]\bar J_{ai} = 0
\eeq
from which, in matrix notation,
\beq
\tilde c=c-(J\bar J^T+\bar J J^T)^{-1}(\bar J f+J \bar f)\;.
\eeq
We see that for this method we should also know the derivatives of our $F_a(u)$ w.r.t. the parameters $c_{a,n}$,
which in our implementation we find numerically by shifting a bit the corresponding parameter.

In some cases, when the starting points are far away from the minimum, the above procedure may start to diverge.
In such cases one can switch to a slower, but more stable, gradient descent method for a few steps. The Levenberg-Marquardt algorithm
provides a nice way to interpolate between the two algorithms by inserting a positive parameter $\Lambda$ into the above
procedure,
\beq
c_{n+1}=c_{n}-(J\bar J^T+\bar J J^T+\Lambda I)^{-1}(\bar J f+J \bar f)\;.
\eeq
The point is that for large $\Lambda$ this is equivalent to the gradient descent method. Thus
one can try to increase $\Lambda$ from its zero value until $S(c_{n+1})<S(c_n)$ and only then
accept the new value $c_{n+1}$. This helps a lot to ensure stable convergence.

In the next section we demonstrate the performance of our numerical method by applying it to the twist-2 operators in $\msl$ sector.

\subsection{Implementation for the $\mathfrak{sl}(2)$ Sector and Comparison with Existing Data}

\paragraph{The $\msl$ sector in the QSC framework.}
\label{sec:Application}
Although our method can be used for any state of the
${\cal N}=4$ SYM theory, the examples we provide in this paper are for states in the $\msl$ subsector. In this section we will discuss the physical operators which have integer spin, and demonstrate our numerical method in action for the Konishi operator. Then in section \ref{sec:nonint} we will show how the algorithm works for other states with $S$ no longer an integer.

Let us sketch the needed information about the QSC in the $\msl$ subsector (more details can be found in \cite{PmuLong,PmuPRL,Alfimov:2014bwa}).
Operators in this sector have only three non-zero quantum numbers: spin $S\equiv S_1$, twist $L\equiv J_1$ and conformal dimension $\Delta$.
Twist-$L$ single-trace operators of this subsector can be schematically represented as
\beq\la{twistL}
{\cal O}=\Tr\(D^S Z^L\)+\dots,
\eeq
where $D$ is a light-cone derivative, $Z$ is a scalar of the theory and the dots stand for permutations.

For such states the QSC enjoys several simplifications. First, quantities with upper and lower indices are now related to each other: indices can be raised or lowered using a simple matrix\footnote{Notice that the choice of $\chi$ fixes the leading order of $Q_{a|i}$ up to a rescaling.}
\beq\label{chi}
\chi=\left(
                                                                                                                                                                                                                  \begin{array}{cccc}
                                                                                                                                                                                                                    0 & 0 & 0 & -1 \\
                                                                                                                                                                                                                    0 & 0 & 1 & 0 \\
                                                                                                                                                                                                                    0 & -1 & 0 & 0 \\
                                                                                                                                                                                                                    1 & 0 & 0 & 0 \\
                                                                                                                                                                                                                  \end{array}
                                                                                                                                                                                                                \right),
\eeq
 for example,

\beq\la{PQchi}
\bQ^i=\chi^{ij}\bQ_j,\; \bP^a=\chi^{ab}\bP_b.
\eeq
It is also easy to show that in this sector $\omega_{ij}$ should satisfy $\omega_{14}=\omega_{23}$ in addition to antisymmetry.
 Second, Q-functions have now definite parity in $u$, which decreases the number of expansion coefficients in series like \eqref{Qlarge} two times. Finally, one can write down a simplified version of asymptotics \eqref{asymptPQ}:
 \beqa
&&\bP_a\sim(A_1u^{-L/2-1},A_2u^{-L/2},A_3u^{L/2-1},A_4u^{L/2}),
\label{eq:asymptoticsP}
\\
&&\bQ_a\sim(B_1u^{\frac{\Delta-S}{2}},B_2u^{\frac{\Delta+S-2}{2}},B_3u^{-\frac{\Delta+S}{2}},B_4u^{\frac{-\Delta+S-2}{2}}).
\label{eq:asymptoticsQ}
\eeqa
Coefficients $A_a$ and $B_i$ are related to the global charges $L,S, \Delta$ (see \cite{PmuPRL,PmuLong}).

\FIGURE[ht]{
\includegraphics[scale=0.8]{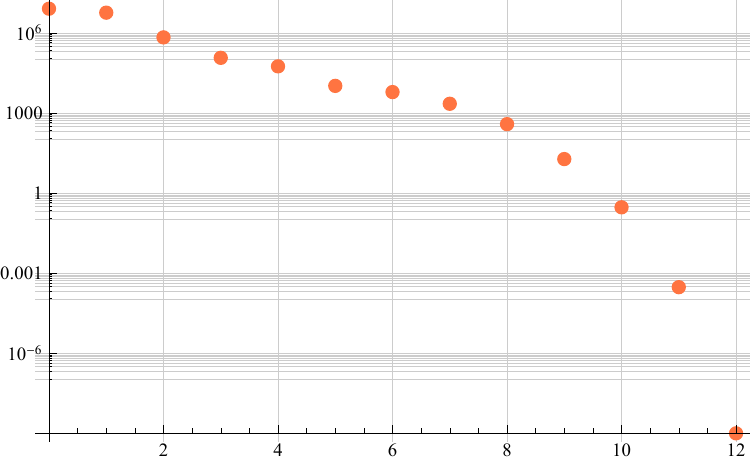}
\caption[a]{{\textbf{Convergence of the algorithm.}} The error $\epsilon_n$ as measured by the value of \eq{Sdef} reduces at the quadratic rate $\epsilon_n\sim e^{-c\; 2^n}$
as a function of the iteration number. In most cases our program managed to find the solution from a very remote starting point.
On the picture we started from all free parameters $c_{a,n}$ set to zero and with the initial value for the energy
$\Delta_0=4.1$.
After $12$ iterations it correctly reproduced $\Delta=4.4188599$
at $\lambda = 16\pi^2 (0.2)^2\simeq 31.6$. With each iteration taking about $1.5$sec
the whole procedure took about  $20$ sec on a Laptop with Intel i7 2.7GHz processor.
}
\la{fig:conv}
}

\paragraph{Improved implementation: skipping the computation of $\omega$'s.}

We have mentioned before that the simplification of the QSC achieved in \cite{Gromov:2015vua} should allow to significantly improve the iterative procedure, as one can avoid calculating $\omega_{ij}$. Here we present this improvement for symmetric states in the $sl(2)$ sector. Let us briefly recall the trick used in \cite{Gromov:2015vua} to eliminate $\omega$'s. For the states we consider, each of the $\bP_a(u)$ functions is either even or odd. Then, as follows from the 4th order finite difference eqation on $\bQ_i$ with coefficients built from $\bP$'s \footnote{its explicit form is given e.g. in (3.2) of \cite{Alfimov:2014bwa}}, $\bQ_i(-u)$ satisfy the same finite difference equation as $\bQ_i(u)$. Thus each of the former can be expressed as a linear combination of the latter with periodic coefficients:
\beq
\label{Qmu}
\bQ_i(u)=\Omega_i^j(u)\bQ_j(-u)\ .
\eeq
We work in the basis where $\bQ_i$ have pure power-like asymptotics at large $u$, non-coinciding for general values of global charges. It is easy to see that at large $u$ in this basis  $\Omega_i^j(u)$ should be constant and diagonal. At the same time, \eq{Qmu} allows us to relate $\bQ_i(-u)$ and $\tilde \bQ_i$:
\beq
\tilde\bQ_i(u)=\alpha_i^j\bQ_j(-u),\;\;\alpha_i^j=\omega_{il}\chi^{lk}\Omega_k^j,
\eeq
where $\chi$ is defined in \eqref{chi}.
The functions $\bQ_i(-u)$ and $\tilde\bQ_i(u)$ have the same analytical properties, so $\alpha_{j}^i$ should be $i$-periodic and analytic. One should also take into account that only $\omega_{12}$ and $\omega_{34}$ are non-zero at infinity, thus many elements of $\alpha_i^j$ have to be zero. For indices $1$ and $3$ we finally get the key relations which appear to be sufficient to close the QSC equations:
\beq
\tilde\bQ_1(u)=\alpha_{13}\bQ_3(-u),\;\;\tilde\bQ_3(u)=\alpha_{31}\bQ_1(-u)
\eeq
Consistency of these two equations also implies that $\alpha_{13}=1/\alpha_{31}\equiv \alpha$.
Note that $\tilde\bQ_1(u)$ can be constructed in our algorithm as $\Q_{a|1}^+\tilde\bP^a$. The equation above tells us that it should be proportional to $\bQ_3(-u)$ with unknown constant of proportionality. This requirement can be also phrased as a minimization problem. For that let us evaluate the ratio $\tilde\bQ_1(u)/\bQ_3(-u)$ at sampling points $u_k$ on the cut $[-2g,2g]$ and compute its variance, 
\beq
	S=\sum_{k=1}^M \left|\frac{\Q_{a|1}(u_k+i/2)\tilde\bP^a(u_k)}{\bQ_3(-u_k)}-B\right|^2\ ,
\eeq
where the constant $B$ is the mean value,
\beq
	B=\frac{1}{M}\sum_{k=1}^M\frac{\Q_{a|1}(u_k+i/2)\tilde\bP^a(u_k)}{\bQ_3(-u_k)}\ .
\eeq
On the true solution of the QSC this ratio is a constant so the variance should be zero, i.e. $S=0$. Thus our goal is to minimize the function $S$, and for this we again use the Levenberg-Marquardt algorithm described above. This gives the desired numerical prediction for the coefficients $c_{a,n}$ parameterizing the $\bP$-functions. 

%
The main performance gain stems from the fact that as we do not compute $\omega$'s, we no longer need to calculate the integrals \eqref{int} and \eqref{alphaI}. This improved method can also be used for non left-right symmetric states as well\footnote{It has been applied to study the BFKL regime with nonzero conformal spin in \cite{ToAppearBFKLN}}, with simple modifications (details of this generalization will be presented elsewhere).

\paragraph{Implementation for Konishi}
Here we discuss the convergence on a particular example of the Konishi operator which corresponds to $S=2,L=2$.
The reason we start from this operator is because it is very well studied both analytically at weak and strong coupling and also numerically.
So we will have lots of data to compare with.

To start the iteration process described in the previous sections, we need some reasonably good starting points for the coefficients $c_{a,n}$.
For the iterative methods, like, for instance, Newton's method, good starting points are normally very important. Depending on them the procedure may converge very slowly
or even diverge. We made a rather radical test of the convergence of our method by setting all coefficients to zero, except the leading ones, which are fixed by the
charges. For $\Delta$ we took the initial value $4.1$ at the value of 't Hooft coupling $g=0.2$.
To our great surprise it took only $12$ steps to converge from the huge value of $S(c_a)\sim 10^{+7}$ (defined in \eq{Sdef})
to $S(c_a)\sim 10^{-9}$. The whole process took about $20$ seconds on a usual laptop (see Fig.~\ref{fig:conv}),
producing the value $\Delta=4.4188599$, consistent with the best TBA estimates \cite{Gromov:2009zb,Frolov:2010wt}.

After that we used the obtained coefficients as starting points for other values of the coupling to produce a large volume of data, part of which is shown in Table \ref{tab:Konishi}.
All the values obtained are consistent with the TBA results within the precision of the latter, being considerably more accurate at the same time.


\TABLE{
\la{tab:Konishi}

\begin{tabular}{|l|l||l|l|}
\hline
$\frac{\sqrt\lambda}{4\pi}$ & $\Delta_{S=2}(\lambda)$ &
$\frac{\sqrt\lambda}{4\pi}$ & $\Delta_{S=2}(\lambda)$
\\ \hline
 0.1 & 4.115506377945221056840042671851 & 0.2 & 4.418859880802350962250362876243 \\
 0.3 & 4.826948662284842304671283425271 & 0.4 & 5.271565182595898008221528540034 \\
 0.5 & 5.712723424787739030626966875973 & 0.6 & 6.133862814488691819595425762346 \\
 0.7 & 6.531606077852440195886557953690 & 0.8 & 6.907504206024567515828872789717 \\
 0.9 & 7.2641695874391127748396398539 & 1 & 7.60407071704738848334286555 \\
 1.1 & 7.9292942641568451632186264 & 1.2 & 8.241563441147703542676050 \\
 1.3 & 8.54230287229506674486342 & 1.4 & 8.8326999393163090494514 \\
 1.5 & 9.11375404891588560886 & 1.6 & 9.386314656368554140399 \\
 1.7 & 9.65111042653013781471 & 1.8 & 9.9087717085593508789 \\
 1.9 & 10.1598480131615473641 & 2 & 10.4048217434405061127 \\
 2.1 & 10.6441190951617575972 & 2.2 & 10.878118797537726796 \\
 2.3 & 11.107159189584305149 & 2.4 & 11.331544000504529107 \\
 2.5 & 11.551547111042160297 & 2.6 & 11.76741650605722239 \\
 2.7 & 11.97937757952067741 & 2.8 & 12.18763591669137588 \\
 2.9 & 12.3923796509149519 & 3 & 12.5937814717988565 \\
 3.1 & 12.7920003457144898 & 3.2 & 12.9871829973986392 \\
 3.3 & 13.1794651919629055 & 3.4 & 13.368972849208144 \\
 3.5 & 13.555823016292914 & 3.6 & 13.740124720157966 \\
 3.7 & 13.921979717391474 & 3.8 & 14.101483156227149 \\
 3.9 & 14.278724162943763 & 4 & 14.45378636296056 \\
 4.1 & 14.62674834530641 & 4.2 & 14.79768407780976 \\
 4.3 & 14.96666327925592 & 4.4 & 15.13375175384302 \\
 4.5 & 15.29901169250472 & 4.6 & 15.4625019450274 \\
 4.7 & 15.6242782663505 & 4.8 & 15.7843935399844 \\
 4.9 & 15.942897981092 & 5 & 16.099839321454 \\ \hline
\end{tabular}

\caption{Conformal dimension of Konishi operator}
}

The reason for such an excellent convergence is the Q-quadratic convergence rate of the algorithm we use. It means that the number of exact digits doubles with each iteration,
or that the error decreases as $e^{-c\; 2^n}$ at the step $n$, if the starting point is close enough. What is perhaps surprising is that the algorithm converges
from a very remote starting point.

Another indicator of the convergence is the plot of $\tilde\bQ$ computed in two different ways, i.e. \eq{QfromQait} and \eq{muab1}.
On the true solution of the QSC both should coincide. On Fig. \ref{fig:Qs} we show how fast the difference between them vanishes with iterations,
i.e. how fast we approach the exact solution of the QSC.

Our numerical data allows to verify with high precision the strong coupling analytic prediction for the Konishi dimension obtained in \cite{Roiban:2009aa,Gromov:2009zb,Gromov:2011de,Roiban:2011fe,Vallilo:2011fj,Gromov:2011bz,Frolov:2013lva,Gromov:2014bva},
\beqa
\label{KonStr}
	\Delta_{konishi}^{an}&=&2\,\lambda^{1/4}+\frac{2}{\lambda^{1/4}}+\frac{-3\,\zeta_3+\ofrac{2}}{\lambda^{3/4}}+\frac{\frac{15 \, \zeta_5}{2} + 6 \, \zeta_3+\frac{1}{2}}{\lambda^{5/4}}+\dots\ 
	\\ \nn
	&=&2 \lambda^{1/4}+\frac{2}{\lambda^{1/4}}
	-
   \frac{3.106170709}{\lambda^{3/4} }+\frac{15.48929958 }{\lambda^{5/4}
   }+\dots
\eeqa
Fitting our data we find
\beq
\label{KonStrN}
	\Delta_{konishi}^{num}-\Delta_{konishi}^{an}=-{0.0000000475}\lambda^{1/4}
	+\frac{0.0000277}{\lambda^{1/4}}+\frac{0.0075}{\lambda^{3/4}}+\frac{1.265}{\lambda^{5/4}}+\dots
\eeq
where we see that the first coefficient is reproduced with high precision. Fixing its value to match the analytic result, we find that the precision of the fit at subsequent orders in $1/\lambda$ improves. Proceeding in this way, we have confirmed all coefficients in the analytic prediction \eq{KonStr} with at least 3-4 digits precision, which is far more than the TBA data allows to get at these high orders in the expansion.




In the next section we discuss the analytic continuation in $S$
away from its integer values. This is an important calculation which bring us to a highly accurate numerical estimate for the pomeron intercept at finite coupling ~---
a quantity which can be studied exclusively by our methods.

\FIGURE[ht]{
\la{fig:Qs}
\includegraphics[scale=0.88]{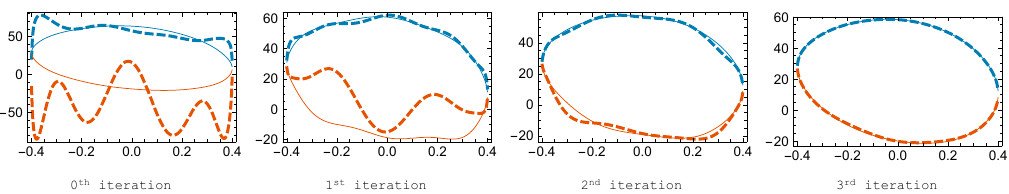}
\caption{\textbf{Q-functions at the first several iterations.}
 Here we show how $\bQ_3$ converges to the solution in just four iterations when calculating the Konishi anomalous dimension. At each picture solid and dashed blue lines show $\bQ_3$ slightly below the cut calculated with {\protect\eqref{muab1}} and {\protect\eqref{QfromQait}} respectively, which should coincide on the solution. Red lines show the same slightly above the cut.
}
}

\section{Extension to Non-Integer Lorentz Spin}
\label{sec:nonint}
In this section we explain which modifications are needed in order to extend our method to non-integer values of spin $S$, and give two specific examples of calculations for such $S$.

\subsection{Modification of the Algorithm for Non-Integer Spin}
\la{sec:Zero}
First we need to discuss how the procedure of fixing zero modes of $\omega$'s described in section \ref{sec:Recovering} is modified for non-integer $S$. The main difference stems from the fact that analytic continuation to non-integer $S$ changes the asymptotic behavior of $\omega_{ij}$ at large $u$, as described in \cite{Gromov:2014bva,Alfimov:2014bwa,ABA}. While for integer $S$ asymptotics of $\omega$ are constant, for non-integer $S$ some components of $\omega$ have to grow exponentially. Without this modification the system has no solution: indeed, in section \ref{sec:Finding} we assumed constant asymptotics of all $\omega$'s and derived quantization condition for global charges.

A minimal modification would be to allow exponential asymptotics in one of the components of $\omega$. In order to understand which of the components can it be, let us recall the Pfaffian constraint satisfied by $\omega_{ij}$
\beq
   \textrm{Pf}\;\omega=\omega_{12}\omega_{34}-\omega_{13}\omega_{24}+\omega_{14}^2=1.
   \la{pff}
\eeq
First, it is clear $\omega_{14}$ alone can not have exponential asymptotics.
Second, in the case of integer $S$ both $\omega_{12}$ of $\omega_{34}$ are non-zero constants at infinity \cite{Alfimov:2014bwa,Gromov:2014bva}; then shifting $S$ infinitesimally away from an integer we see that it would be impossible to satisfy the condition \eq{pff} if we allow one of them to have exponential asymptotics at infinity: this exponent will multiply the constant in the other one. So the only two possibilities left are $\omega_{13}$ and $\omega_{24}$, which are both zeros at infinity for integer $S$. From perturbative data we know that it is $\omega_{24}$ which should have exponential asymptotics. Thus we formulate the ``minimal'' prescription for analytic continuation of $Q$-system to non-integer $S$: $e^{2\pi |u|}$ asymptotic has to be allowed in $\omega_{24}$. This prescription was tested thoroughly on a variety of examples \cite{Janik:2013nqa,GrKazakov,Alfimov:2014bwa,Gromov:2014bva,ABA}, but it would be interesting to derive it rigorously and generalize it to states outside of the $\msl$ sector. Of course, one can also consider adding exponents to more than one component of $\omega_{ij}$: in this case the solution will not be unique.
A complete classification of solutions of
Q-system according to exponents in their asymptotics might be interesting.
For example it is known that allowing for an exponent in some other components corresponds to the generalized cusp anomalous dimension \cite{Gromov:2014bva,CuspPmu}.

Because of the exponential asymptotics of $\omega_{24}$, the argument in section \ref{sec:Recovering}, which fixes the zero modes of $\omega$, has to be modified. First, formula \eq{omegaI} still holds true for $i=1$ or $i=3$, as $\omega_{24}$ does not enter anywhere in the derivation. Thus
\beq
\omega_{12}=-i I_{12}\cot \frac{\pi\(S+\Delta\)}{2},\;\omega_{34}=-i I_{34}\cot \frac{\pi\(S-\Delta\)}{2}.
\eeq
Consequently, one can use \eq{omegaI} for both $\omega_{13}$ and $\omega_{31}$, and reproduce the quantization condition \eq{antisymm} for global charges, which in this case implies that either $\Delta=0$ or $\omega_{13}=0$.
Equation \eq{omegaI} can also be used for $\omega_{14}$ and $\omega_{23}$ (which are equal) and imposes that either $\Delta=0$ or $\omega_{14}=0$.

 It remains to fix the zero mode in $\omega_{24}^0$, for which we use an ansatz
\beq
\omega_{24}^0=a_1 e^{2\pi u}+ a_2+ a_3 e^{-2\pi u}.
\eeq
The constants $a_1,a_2,a_3$ can be found from the Pfaffian constraint \eq{pff}.
To this end we expand the hyperbolic cotangent in \eq{int} to get
\beq
\omega_{ij}=\omega_{ij}^0+I_{ij}+2e^{-2\pi u}I^+_{ij}+2e^{-4\pi u}I^{++}_{ij}+\dots,\; u\rightarrow+\infty,
\eeq
where the terms of the expansion are integrals similar to $I_{ij}$ with additional factors of $e^{2\pi u}$ or $e^{4\pi u}$ in the integrand\footnote{Actually, these integrals can be evaluated analytically in terms of Bessel functions}. Analogous expansion can be obtained at $u\rightarrow -\infty$.
Then plugging these expansions into \eq{pff} we get formulas for the coefficients $a_1,a_2,a_3$. For example,
\beq\la{zero1}
a_1=2i\frac{1+ \frac{I_{12}I_{34}}{4}\(1+i\cot\frac{\pi(\Delta+S)}{2}\)\(1-i\cot\frac{\pi(\Delta-S)}{2}\)  }{I_{13}^+}\;.
\eeq

With these modifications we can reconstruct all $\omega_{ij}$ including the zero modes and then proceed with our algorithm as in the case of integer $S$.

\subsection{Exploring Complex Spin}

\FIGURE[ht]{
\includegraphics[scale=0.6]{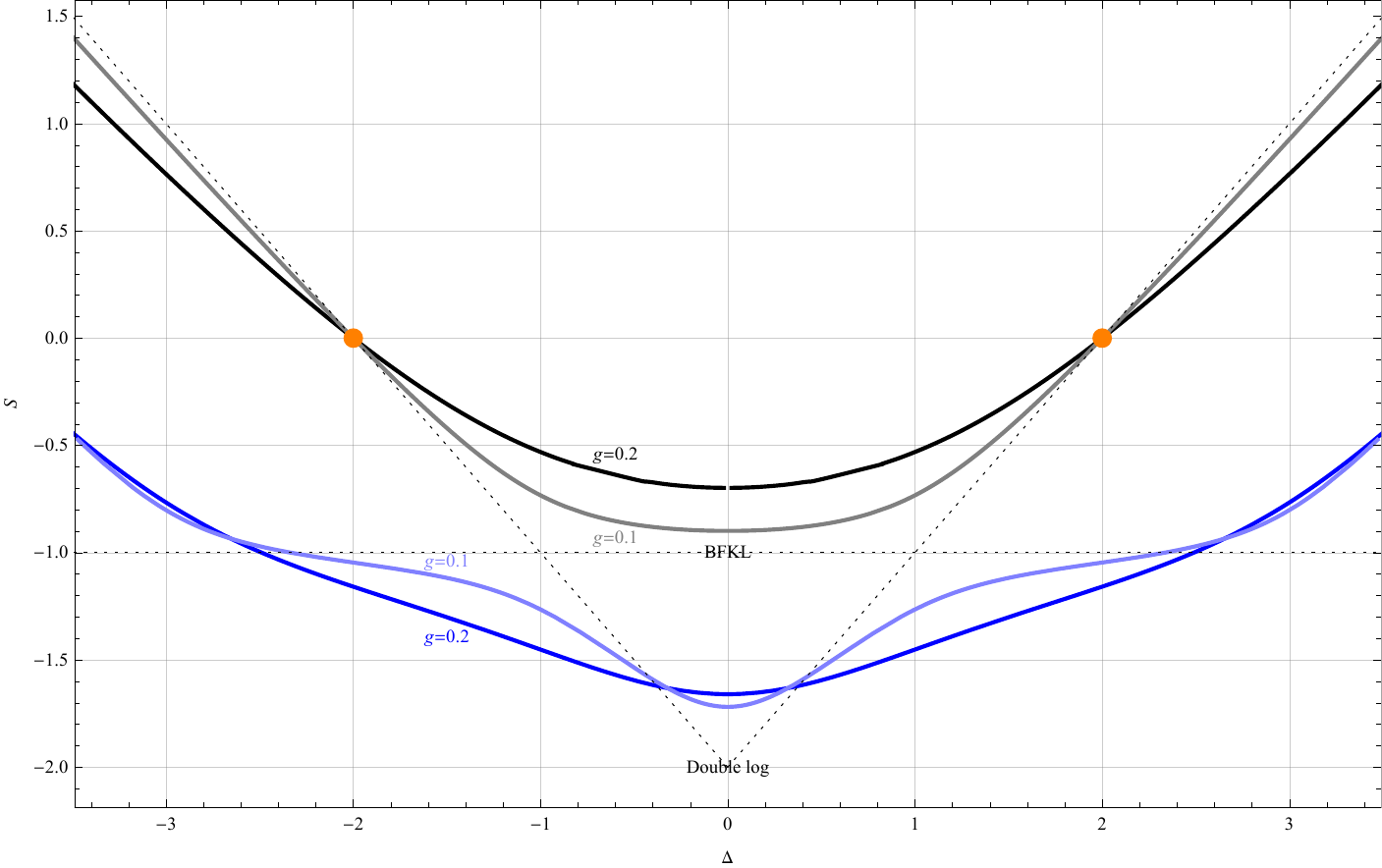}
\caption{Section of the Riemann surface $S(\Delta)$ along ${\rm Im} \; \Delta=0$ for different values of coupling $g$. The upper two solid curves, shown in black and grey, represent the well-known BFKL eigenvalue as a function of $\Delta$, whereas the lower two come from the unphysical sheet which can be accessed from the upper one by going through the cuts. The dashed line shows the zero-coupling limit of the curve. Orange dots mark BPS states $\Tr(ZZ)$.}
\la{fig:sec}
}

In this section we briefly describe the results of our numerical exploration of
$\Delta(S)$ as an analytic function of a complexified spin $S$.
As explained in the previous section the generalization of our numerical method to
arbitrary values of spin requires minimal modifications of our main code.
Thus we are able to generate numerous
values of the anomalous dimension for any $S$ with high precision in seconds.
In fact both $S$ and $\Delta$ enter the QSC formalism on almost equal footing and
we can also switch quite easily to finding $S$ for given $\Delta$.
This is what is adopted in the vast literature on the subject and what we are going to consider below.
This viewpoint is particular convenient due to the symmetry $\Delta\to -\Delta$
which makes the pictures particularly nice.

Starting from $S=2$ (Konishi operator) we decreased the value of $S$ or $\Delta$ in small steps using
the solution at the previous step as a starting point for the next value.
In this way we built the upper two curves on Fig.~\ref{fig:sec}.
Let us point out one curious technical problem -- one can see for instance from \eq{zero1}
that the lines $S=\pm \Delta+ {\mathbb Z}$ are potentially dangerous due to the divergence.
In fact we found that near these dangerous points on the line the factor $I_{12}I_{34}$
also vanishes canceling the potential divergence. This however affect the convergence ``radius"
of our iterative procedure and we found it quite complicated to cross those lines, even though in
very small steps we were able to reach close to them.
The way out is to go around these lines in the complex plane $\Delta$.
To make sure there is no true singularity or branch point we also explored a big patch of the complex
plane $\Delta$, indeed finding some branch points, but deep inside the complex plane, having nothing to do
with these lines. For example when $g=0.2$ we found $4$ closest branch points at roughly $\pm 1\pm i$,
see Fig.~\ref{fig:sdelta}. By making an analytic continuation (described above) through those cuts we found
another sheet of the Riemann surface $S(\Delta)$.
On this sheet we have found four cuts: two are connecting it to
the first sheet and two other ones, located symmetrically on the imaginary axis, lead to
further sheets. We expect an infinite set
of sheets hidden below and also more cuts on both sheets outside of the area that we have explored.

\FIGURE[ht]{
\la{fig:intercept}
\includegraphics[scale=1]{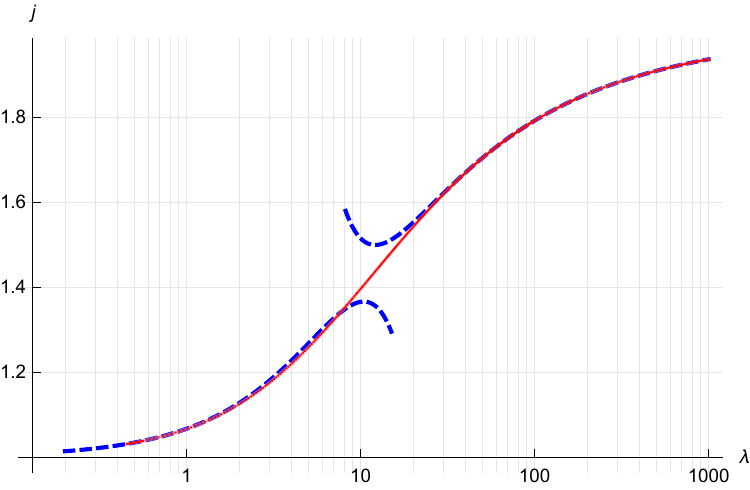}
\caption{The BFKL intercept $j$ as a function of coupling $\lambda$. The red solid line with tiny red dots is obtained by our numerical procedure.
It interpolates perfectly between the known perturbative predictions
(the blue dashed lines)
 at weak \cite{Kotikov:2002ab,Brower:2006ea} and strong coupling \cite{Costa:2012cb,Kotikov:2013xu,Brower:2013jga,Gromov:2014bva}.}
}

It is instructive to see how this Riemann surface behaves as $g\rightarrow 0$.
First, the real parts of branch points on the physical sheet are very close to $\pm1$, but the imaginary part goes to zero. Thus at infinitely small $g$ the cuts collide, isolating the region $|\Re\; \Delta|<1$ from the rest of the complex plane. These two separated regions become then the areas of applicability of two different approximations: for $|\Re\; \Delta|>1$ one can apply the usual perturbation theory and Beisert-Eden-Staudacher Asymptotic Bethe Ansatz, whereas the region $|\Re\; \Delta|<1$ is described by BFKL approximation and so-called Asymptotic BFKL Ansatz \cite{ABA}.

The presence of the cut can be to some extent deduced from perturbative perspective in each region: in the regime of usual perturbation theory
\beq
\Delta(S)=2+S-8g^2 H_S+\cO(g^4),
\eeq
where $H_S$ is the harmonic number. It has poles for all negative integer values of $S$ ~--- these poles are weak-coupling remnants of the cuts we see at finite coupling. In the BFKL regime one should instead look at the leading order BFKL equation \cite{Jaroszewicz:1982gr,Lipatov:1985uk,Kotikov:2002ab}
\beq
S(\Delta)=-1+4g^2\[\psi\(\frac{1+\Delta}{2}\)+\psi\(\frac{1-\Delta}{2}\)-2\psi(1)\]+{\cal O}(g^4)\;.
\eeq
To make sense of this equation one has to take the limit $g\rightarrow 0$, $S\rightarrow-1$ so that the l.h.s stays finite. Then the $\psi$-functions in the r.h.s generate poles at odd values of $\Delta$, which, again, are cuts degenerated at weak coupling.

Fig \ref{fig:sec} represents a section of the Riemann surface by the plane $\Im\;u=0$, i.e. dependence of $S$ on $\Delta$ for real $\Delta$, which, of course, consists of two curves, originating from the two sheets we explored. At weak coupling the upper curve becomes piecewise linear, approaching different parts of the dotted line: for $|\Delta|>1$ it coincides with $S=\pm\Delta-2$ and for $|\Delta|<1$ it becomes $S=-1$. One could expect a similar piecewise linear behavior for the lower curve: it approaches $S=\pm\Delta-2$ for $|\Delta|<1$, approaches $S=0$ in some region outside of $|\Delta|<1$ and becomes a certain linear function even further away from $\Delta=0$.
It would be interesting to explore the complete analytic structure of this Riemann surface, and
understand what describes its asymptotics when $g\to 0$. It should produce a hierarchy of ``Asymptotic Bethe Ans\"atze"
each responsible for its own linear part of the limiting surface.

\subsection{BFKL Pomeron Intercept}

The pomeron intercept $j(\lambda)$ is a quantity which relates spectrum of single-trace operators and scaling of high energy scattering amplitudes in the Regge regime \cite{bfkl}. This regime is particularly interesting, since it establishes a connection between results in ${\cal N}=4$ SYM and multicolor QCD: the non-trivial leading order of so-called BFKL eigenvalue is the same in two theories, and in the higher orders ${\cal N}=4$ SYM is expected to reproduce at least the most complicated part of the QCD result.

Our goal is to demonstrate the universal power of our approach by giving an extremely precise numerical estimate for this
important quantity at finite coupling in a wide range of couplings.

One defines the intercept as $j=S(\Delta=0)+2$, where $S$ is the spin of the twist-2 operator such that $\Delta(S)=0$.
Having formulated the problem like this, we can in principle apply
the algorithm described in section \ref{sec:Solving} to find the correct value of $S$,
while keeping $\Delta$ at zero.
However, one may already suspect that the point $\Delta=0$ is special.
Indeed, we know that for any solution of QSC there is always another one related by $\Delta \to-\Delta$ symmetry.
At the level of $\bQ_i$ functions this allows simultaneously interchanging $\bQ_1\leftrightarrow \bQ_3$ and
$\bQ_2\leftrightarrow \bQ_4$ as one can see from the asymptotics \eqref{eq:asymptoticsQ}.
From this we see that at small $\Delta$ two different solutions of QSC (related by the symmetry)
approach each other, making the convergence slow, exactly like Newton's method becomes inefficient for degenerate zeros. In other words, in the limit $\Delta\to 0$ the $\bQ$'s related by the symmetry
become linearly dependent in the leading order. Furthermore, since the matrix ${\cal Q}_{a|i}$ should stay invertible, the leading coefficients $B_i$ of asymptotic expansion of $\bQ_i$ diverge at $\Delta\rightarrow 0$.

The way out is to perform a linear transformation of $\bQ$'s preserving the equations: it will replace two of them by linear combinations $\bQ_3-\gamma\bQ_1$ and
$\bQ_4+\gamma\bQ_2$ with some coefficient $\gamma$, so that the divergent leading order cancels and the four functions $\bQ_i$ become linearly independent.

For the gauge choice in which $B_1=B_2=1$ the transformation acts on $i$-indices of Q-functions with a matrix\footnote{This is a particular case of $H$-transformations described in section 4.1.3 of \cite{PmuLong}}.
\beq
H_i{}^j=\left(
  \begin{array}{cccc}
    1 & 0 & 0 & 0 \\
    0 & 1 & 0 & 0 \\
    -\gamma & 0 & 1 & 0 \\
    0 & \gamma & 0 & 1 \\
  \end{array}
\right),\; \gamma=\frac{i(S-4)(S-2)S(S+2)}{16(S-1)^2\Delta}.
\eeq

One can check that rotation by this matrix will render ${\cal Q}_{a|i}$ finite and linearly independent, and moreover, preserve relations \eqref{PQchi}. After this one can apply the standard procedure from section \ref{sec:Solving}
with the only modification that the large $u$ expansion of ${\cal Q}_{a|i}$  will contain $\log u/u^n$ terms in addition to the usual $1/u^n$.

Having done this, we can readily generate lots of numerical results. In particular
we built numerically the function $j(\lambda)$ which interpolates perfectly between the weak and
strong coupling predictions. We have found
$j(\lambda)$ with high precision (up to $20$ digits) for a wide range of 't Hooft coupling
(going up to $\lambda\simeq 1000$). The results are also summarized in the Table \ref{tab:intercept}. This table represents a small portion of all data we generated, which is available by request.

\TABLE{
\begin{tabular}{|l|l||l|l|}
\hline
$\frac{\sqrt\lambda}{4\pi}$ & $j(\lambda)$ &
$\frac{\sqrt\lambda}{4\pi}$ & $j(\lambda)$
\\ \hline
 $0.$ & ${1.000\,000\,000\,000\,000\,000\,0}$ & $0.1$ & $ {1.101\,144\,978\,997\,772\,874\,8}$
  \\
 $0.2$ & $ {1.301\,794\,032\,258\,782\,208\,7}$
   & $0.3$ &
   $ {1.470\,445\,240\,989\,187\,630\,6}$
\\
 $0.4$ & $ {1.587\,128\,066\,254\,129\,730\,4}$
 & $0.5$ & $ {1.666\,438\,709\,974\,061\,852\,3}$
\\
 $0.6$ & $ {1.721\,917\,842\,815\,631\,353\,9}$
   & $0.7$ &
   $ {1.762\,239\,296\,816\,453\,814\,3}$
\\
 $0.8$ & $ {1.792\,626\,253\,069\,403\,59}\hid{4\,0}$
    & $0.9$ &
   $ {1.816\,252\,952\,807\,284\,11}\hid{6\,4}$
\\
 $1.$ &
              $ {1.835\,109\,464\,032\,173\,0
              }\hid{85\,4}$
  & $1.1$ &
$    {1.850\,489\,553\,739\,522\,8}\hid{493}$
\\
$ 1.2$ & $ {1.863\,264\,346\,392\,640\,4
}\hid{78\,5}                                  $
  & $1.3$ &
$    {1.874\,039\,320\,799\,460}\hid{0735}$
\\
$ 1.4$ & $ {1.883\,247\,290\,966\,33}\hid{1\,855\,8}$
   & $1.5$ &
$    {1.891\,205\,346\,040\,23}\hid{0\,357\,1}$
   \\
$ 1.6$ & $ {1.898\,150\,851\,852\,49
 }\hid{7\,559\,2}$
 & $1.7$ &
$    {1.904\,264\,892\,928\,17}\hid{1\,250\,0}$
\\
$ 1.8$ &
$  {1.909\,687\,948\,271\,74
 }\hid{6\,858\,7}$
 & $1.9$ &
$    {1.914\,530\,628\,017\,38
   }\hid{5\,914\,1}$
\\
$ 2.$ & $ {1.918\,881\,187\,304\,9
 }
 \hid{66\,353\,6}$
 & $2.1$ &
$      {1.922\,810\,887\,750
     }\hid{\,754\,882\,5}$
\\
$ 2.2$ &
$              {1.926\,377\,890\,67}\hid{21139373}$
 & $2.3$ &
 $   {1.929\,630\,129\,41\,
   }\hid{9200585}$
\\
$ 2.4$ & $ {1.932\,607\,459\,1}\hid{123778548}$
 & $2.5$ &
$    {1.935\,343\,287\,2}\hid{168490952}$
\\ \hline
\end{tabular}
\caption{Numerical data for the pomeron intercept for various values of the 't Hooft coupling. \la{tab:intercept}
All digits are expected to be significant.}
}

At strong coupling we can confirm the analytic predictions obtained in \cite{Costa:2012cb,Kotikov:2013xu,Brower:2013jga,Gromov:2014bva},
\beqa
\label{IntercStr}
j^{an}&=&2-\frac{2}{\lambda^{1/2} }-\frac{1}{\lambda }+\frac{1}{4} \frac{1}{\lambda^{3/2} }+\frac{6
   \zeta (3)+2}{\lambda ^2}+\frac{18 \zeta
   (3)+\frac{361}{64}}{\lambda^{5/2} }+\frac{39 \zeta (3)+\frac{511}{32}}{\lambda ^3}+\dots\\ \nn
	&=&2-\frac{2}{\lambda^{1/2} }-\frac{1}{\lambda }+ \frac{0.2500000000}{\lambda^{3/2} }
	+\frac{9.212341419}{\lambda ^2}+ \frac{27.27764926}{\lambda^{5/2} }
	+\frac{62.84896922}{\lambda ^3}+\dots
\eeqa
Fitting our numerical results, we find
\beqa
	&&j^{num}-j^{an}=
	\\ \nn &&-{0.000000687}+\frac{0.000149}{\lambda^{1/2}}+\frac{0.0146}{\lambda}
	+\frac{0.854}{\lambda^{3/2}}-\frac{33.105}{\lambda^2}+\frac{892.72}{\lambda^{5/2}}
	-\frac{17093.5}{\lambda^3}\ .
\eeqa
As for the Konishi anomalous dimension (see \eq{KonStrN}), we see that the leading coefficient is reproduced with high precision, while at the next orders the precision decreases and we get coefficients very different from the analytic prediction. However, fixing the leading coefficient to match the analytic result, we again found that the precision of our fit at the next orders in $1/\lambda$ increases. Gradually reproducing the coefficients in this way, we confirmed all coefficients in the analytic result \eq{IntercStr} with the precision of at least 3-4 digits, except for the $\lambda^3$ term for which we could not get a stable fit with the data we have so far. 

Moreover, we generated $\sim100$ points with small $g$ in the range $0.1\dots 0.017$ each with more than $20$ digits precision.
Fitting this data with powers of $g^2$ we found
\beqa\nn
j=1+11.09035488895912 g^2-84.0785668075 g^4-2543.0481652 g^6+156244.8086 g^8
\eeqa
where the first $3$ terms are known analytically from Feynman diagram  perturbation theory calculations \cite{Kotikov:2002ab,Brower:2006ea}
and their numerical values coincide in all digits with our prediction above.
The last two terms give our numerical prediction for the numerical values of the  NNLO and NNNLO BFKL pomeron intercept.
Our fit also gives predictions for the higher corrections but with a smaller precision.

\section{Conclusions and Future Directions}

In this paper we have demonstrated that in addition to their analytic power, the QSC equations can give highly precise numerical results at finite coupling.
We develop a numerical procedure which applies to generic single trace operators and
as such it is unique in its kind.
Furthermore, the algorithm converges at a remarkably high rate which
gives us access to high numerical precision results -- up to $20$ digits or even more in a few iterations.

The efficiency of our method is demonstrated on the example of $\msl$ sector operators.
We also formulated how to extend our procedure to non-integer quantum numbers.
We studied the twist-2 operators for complex values of the spin discovering a fascinating Riemann surface (see Fig.\ref{fig:sdelta}).
In addition we reformulated our equations to be directly applicable to the BFKL pomeron intercept
and evaluated the intercept $j$ with high precision of up to $20$ significant figures.
By fitting our data we also gave a prediction for the perturbation theory expansion
\beqa
j(\lambda)&=&1+0.07023049277268284\; \lambda-0.00337167607361\; \lambda
   ^2\\ \nn
   &-&0.00064579607573\; \lambda ^3
   + 0.0002512619258\; \lambda ^4+\dots
\eeqa
reproducing correctly the first two nontrivial orders \cite{Kotikov:2002ab,Brower:2006ea} and giving a prediction for higher orders.

The range of possible applications of our method is vast. First, it is not limited solely to the $\msl$ sector of ${\cal N}=4$ SYM, but
is directly applicable to any single trace operators of the theory. It would be interesting to do an explicit example
of a numerical calculation with our algorithm outside of the $\msl$ sector. For example, the wider class of ${\mathfrak {sl}}(2,{\mathbb C})$ operators (identified in
\cite{ABA}), also exhibiting a BFKL regime, could be
a good candidate to begin with.
Second, we expect our method to be applicable for such non-local operators as the
generalized cusp anomalous dimension and quark--anti-quark potential, DD-brane and other boundary problems \cite{Drukker:2012de,Correa:2012hh,Bajnok:2013wsa,Bajnok:2012xc,CuspPmu}.
Third, it may be interesting to generalize our method to ABJM theory
as well as to various integrable deformations of ${\cal N}=4$ SYM theory.

The numerical results could also be helpful for the analytical exploration of the spectrum -- for instance, in such regimes as BFKL
and   at strong coupling, which remains almost unexplored, and various limiting cases of the generalized cusp.
Furthermore, studying numerical results and the behavior of the generated Q-functions in various limits
can reveal new analytically solvable regimes.

To facilitate further applications and development of our method we attach to this paper
a user-friendly version of our code as a {\it Mathematica} notebook. It provides an implementation of our algorithm in the simplest case.
It may be also useful to convert our {\it Mathematica} code into a faster language such as 
C++ or a similar lower-level language \footnote{A C++ implementation was very recently presented in \cite{Hegedus:2016eop}. }. It should not be difficult as our algorithm is
quite simple and only uses some basic matrix operations.

\section*{Acknowledgements}
We thank M.~Alfimov, B.~Basso, S.~Caron Huot, S.~Leurent and especially V.~Kazakov for discussions.
The research leading to these results has received funding from the
People Programme (Marie Curie Actions) of the European Union's Seventh
Framework Programme FP7/2007-2013/ under REA Grant Agreement No 317089.
We wish to thank
STFC for support from Consolidated
grant number ST/J002798/1.
N.G. would like to thank FAPESP grant 2011/11973-4 for funding his visit to ICTP-SAIFR from Month-Month 2015 where part of this work was done.

\appendix


\begin{thebibliography}{99}



\bibitem{PmuPRL}
  N.~Gromov, V.~Kazakov, S.~Leurent and D.~Volin,
  ``Quantum Spectral Curve for Planar $\mathcal{N} =$ Super-Yang-Mills Theory,''
  Phys.\ Rev.\ Lett.\  {\bf 112} (2014) 1,  011602
  [arXiv:1305.1939 [hep-th]].



\bibitem{PmuLong}
 N.~Gromov, V.~Kazakov, S.~Leurent and D.~Volin,
  ``Quantum spectral curve for arbitrary state/operator in AdS$_5$/CFT$_4$,''
  arXiv:1405.4857 [hep-th].



\bibitem{Gromov:2009tv}
  N.~Gromov, V.~Kazakov and P.~Vieira,
  ``Exact Spectrum of Anomalous Dimensions of Planar N=4 Supersymmetric Yang-Mills Theory,''
  Phys.\ Rev.\ Lett.\  {\bf 103} (2009) 131601
  [arXiv:0901.3753 [hep-th]].



\bibitem{Bombardelli:2009ns}
  D.~Bombardelli, D.~Fioravanti and R.~Tateo,
  ``Thermodynamic Bethe Ansatz for planar AdS/CFT: A Proposal,''
  J.\ Phys.\ A {\bf 42} (2009) 375401
  [arXiv:0902.3930 [hep-th]].



\bibitem{Gromov:2009bc}
  N.~Gromov, V.~Kazakov, A.~Kozak and P.~Vieira,
  ``Exact Spectrum of Anomalous Dimensions of Planar N = 4 Supersymmetric Yang-Mills Theory: TBA and excited states,''
  Lett.\ Math.\ Phys.\  {\bf 91} (2010) 265
  [arXiv:0902.4458 [hep-th]].



\bibitem{Arutyunov:2009ur}
  G.~Arutyunov and S.~Frolov,
  ``Thermodynamic Bethe Ansatz for the AdS(5) x S(5) Mirror Model,''
  JHEP {\bf 0905} (2009) 068
  [arXiv:0903.0141 [hep-th]].


\bibitem{Cavaglia:2010nm}
  A.~Cavaglia, D.~Fioravanti and R.~Tateo,
  ``Extended Y-system for the $AdS_5/CFT_4$ correspondence,''
  Nucl.\ Phys.\ B {\bf 843} (2011) 302
  [arXiv:1005.3016 [hep-th]].


\bibitem{Gromov:2011cx}
  N.~Gromov, V.~Kazakov, S.~Leurent and D.~Volin,
  ``Solving the AdS/CFT Y-system,''
  JHEP {\bf 1207} (2012) 023
  [arXiv:1110.0562 [hep-th]].




\bibitem{Balog:2012zt}
  J.~Balog and A.~Hegedus,
  ``Hybrid-NLIE for the AdS/CFT spectral problem,''
  JHEP {\bf 1208} (2012) 022
  [arXiv:1202.3244 [hep-th]].


\bibitem{Suzuki:2011dj}
  R.~Suzuki,
  ``Hybrid NLIE for the Mirror $AdS_5 x S^5$,''
  J.\ Phys.\ A {\bf 44} (2011) 235401
  [arXiv:1101.5165 [hep-th]].



\bibitem{Bajnok:2013wsa}
  Z.~Bajnok, N.~Drukker, A.~Hegedus, R.~I.~Nepomechie, L.~Palla, C.~Sieg and R.~Suzuki,
  ``The spectrum of tachyons in AdS/CFT,''
  arXiv:1312.3900 [hep-th].


\bibitem{Gromov:2014bva}
  N.~Gromov, F.~Levkovich-Maslyuk, G.~Sizov and S.~Valatka,
  ``Quantum spectral curve at work: from small spin to strong coupling in $ \mathcal{N} $ = 4 SYM,''
  JHEP {\bf 1407} (2014) 156
  [arXiv:1402.0871 [hep-th]].



\bibitem{Marboe:2014gma}
  C.~Marboe and D.~Volin,
  ``Quantum spectral curve as a tool for a perturbative quantum field theory,''
  arXiv:1411.4758 [hep-th].



\bibitem{Marboe:2014sya}
  C.~Marboe, V.~Velizhanin and D.~Volin,
  ``Six-loop anomalous dimension of twist-two operators in planar N=4 SYM theory,''
  arXiv:1412.4762 [hep-th].





\bibitem{Alfimov:2014bwa}
  M.~Alfimov, N.~Gromov and V.~Kazakov,
  ``QCD Pomeron from AdS/CFT Quantum Spectral Curve,''
  arXiv:1408.2530 [hep-th].



\bibitem{Cavaglia:2014exa}
  A.~Cavaglià, D.~Fioravanti, N.~Gromov and R.~Tateo,
  ``Quantum Spectral Curve of the $\mathcal N=$ 6 Supersymmetric Chern-Simons Theory,''
  Phys.\ Rev.\ Lett.\  {\bf 113} (2014) 2,  021601
  [arXiv:1403.1859 [hep-th]].


\bibitem{Gromov:2014eha}
  N.~Gromov and G.~Sizov,
  ``Exact Slope and Interpolating Functions in N=6 Supersymmetric Chern-Simons Theory,''
  Phys.\ Rev.\ Lett.\  {\bf 113} (2014) 12,  121601
  [arXiv:1403.1894 [hep-th]].



\bibitem{Gromov:2009zb}
  N.~Gromov, V.~Kazakov and P.~Vieira,
  ``Exact Spectrum of Planar ${\cal N}=4$ Supersymmetric Yang-Mills Theory: Konishi Dimension at Any Coupling,''
  Phys.\ Rev.\ Lett.\  {\bf 104} (2010) 211601
  [arXiv:0906.4240 [hep-th]].




\bibitem{Roiban:2011fe}
  R.~Roiban and A.~A.~Tseytlin,
  ``Semiclassical string computation of strong-coupling corrections to dimensions of operators in Konishi multiplet,''
  Nucl.\ Phys.\ B {\bf 848} (2011) 251
  [arXiv:1102.1209 [hep-th]].



\bibitem{Vallilo:2011fj}
  B.~C.~Vallilo and L.~Mazzucato,
  ``The Konishi multiplet at strong coupling,''
  JHEP {\bf 1112} (2011) 029
  [arXiv:1102.1219 [hep-th]].



\bibitem{Gromov:2011de}
  N.~Gromov, D.~Serban, I.~Shenderovich and D.~Volin,
  ``Quantum folded string and integrability: From finite size effects to Konishi dimension,''
  JHEP {\bf 1108} (2011) 046
  [arXiv:1102.1040 [hep-th]].



\bibitem{Basso:2011rs}
  B.~Basso,
  ``An exact slope for AdS/CFT,''
  [arXiv:1109.3154 [hep-th]].



\bibitem{Gromov:2011bz}
  N.~Gromov and S.~Valatka,
  ``Deeper Look into Short Strings,''
  JHEP {\bf 1203} (2012) 058
  [arXiv:1109.6305 [hep-th]].





\bibitem{Frolov:2010wt}
  S.~Frolov,
  ``Konishi operator at intermediate coupling,''
  J.\ Phys.\ A {\bf 44} (2011) 065401
  [arXiv:1006.5032 [hep-th]].



\bibitem{Frolov:2013lva}
  S.~Frolov, M.~Heinze, G.~Jorjadze and J.~Plefka,
  ``Static Gauge and Energy Spectrum of Single-mode Strings in AdS5xS5,''
  arXiv:1310.5052 [hep-th].




\bibitem{Gromov:2015vua}
  N.~Gromov, F.~Levkovich-Maslyuk and G.~Sizov,
  ``Pomeron Eigenvalue at Three Loops in $\mathcal N=$ 4 Supersymmetric Yang-Mills Theory,''
  Phys.\ Rev.\ Lett.\  {\bf 115} (2015) no.25,  251601
  doi:10.1103/PhysRevLett.115.251601
  [arXiv:1507.04010 [hep-th]].

\bibitem{ToAppearBFKLN}
M.~Alfimov, N.~Gromov and G.~Sizov, to appear

\bibitem{Roiban:2009aa}
  R.~Roiban and A.~A.~Tseytlin,
  ``Quantum strings in AdS(5) x S**5: Strong-coupling corrections to dimension of Konishi operator,''
  JHEP {\bf 0911} (2009) 013
  doi:10.1088/1126-6708/2009/11/013
  [arXiv:0906.4294 [hep-th]].

\bibitem{Zoubos:2010kh}
  K.~Zoubos,
  ``Review of AdS/CFT Integrability, Chapter IV.2: Deformations, Orbifolds and Open Boundaries,''
  Lett.\ Math.\ Phys.\  {\bf 99} (2012) 375
  [arXiv:1012.3998 [hep-th]].


\bibitem{ABA}
N.~Gromov and G.~Sizov, to appear


\bibitem{Janik:2013nqa}
  R.~A.~Janik,
  ``Twist-two operators and the BFKL regime - nonstandard solutions of the Baxter equation,''
  JHEP {\bf 1311} (2013) 153
  [arXiv:1309.2844 [hep-th]].



\bibitem{GrKazakov}
N. Gromov, V. Kazakov, unpublished (2013)



\bibitem{CuspPmu}
  N.~Gromov and F.~Levkovich-Maslyuk,
  ``Quantum Spectral Curve for a Cusped Wilson Line in N=4 SYM,''
  arXiv:1510.02098 [hep-th].


\bibitem{Kotikov:2002ab}
  A.~V.~Kotikov and L.~N.~Lipatov,
  ``DGLAP and BFKL equations in the N=4 supersymmetric gauge theory,''
  Nucl.\ Phys.\ B {\bf 661}, 19 (2003)
  [Erratum-ibid.\ B {\bf 685}, 405 (2004)]
  [hep-ph/0208220].


\bibitem{Brower:2006ea}
  R.~C.~Brower, J.~Polchinski, M.~J.~Strassler and C.~-ITan,
  ``The Pomeron and gauge/string duality,''
  JHEP {\bf 0712}, 005 (2007)
  [hep-th/0603115].




\bibitem{Costa:2012cb}
  M.~S.~Costa, V.~Goncalves and J.~Penedones,
  ``Conformal Regge theory,''
  JHEP {\bf 1212}, 091 (2012)
  [arXiv:1209.4355 [hep-th]].


\bibitem{Kotikov:2013xu}
  A.~V.~Kotikov and L.~N.~Lipatov,
  ``Pomeron in the N=4 supersymmetric gauge model at strong couplings,''
  Nucl.\ Phys.\ B {\bf 874}, 889 (2013)
  [arXiv:1301.0882 [hep-th]].



\bibitem{Brower:2013jga}
  R.~C.~Brower, M.~Costa, M.~Djuric, T.~Raben and C.~-ITan,
  ``Conformal Pomeron and Odderon in Strong Coupling,''
  arXiv:1312.1419 [hep-ph].




\bibitem{Jaroszewicz:1982gr}
  T.~Jaroszewicz,
  ``Gluonic Regge Singularities and Anomalous Dimensions in QCD,''
  Phys.\ Lett.\ B {\bf 116} (1982) 291.


\bibitem{Lipatov:1985uk}
  L.~N.~Lipatov,
  ``The Bare Pomeron in Quantum Chromodynamics,''
  Sov.\ Phys.\ JETP {\bf 63} (1986) 904
   [Zh.\ Eksp.\ Teor.\ Fiz.\  {\bf 90} (1986) 1536].



\bibitem{bfkl}
  L.~N.~Lipatov, ``Reggeization of the vector meson and the vacuum singularity in nonabelian gauge theories,'' Sov.\ J.\ Nucl.\ Phys.\ {\bf 23} (1976) 338 [Yad. Fiz. {\bf 23} (1976) 642]. $\bullet$
  E.~A.~Kuraev, L.~N.~Lipatov and V.~S.~Fadin, ``The Pomeranchuk singularity in nonabelian gauge theories,'' Sov.\ Phys.\ JETP\ {\bf 45} (1977) 199 [Zh. Eksp. Teor. Fiz. {\bf 72} (1977) 377]. $\bullet$
  I.~I.~Balitsky and L.~N.~Lipatov, ``The Pomeranchuk singularity in Quantum Chromodynamics,\\ Sov.\ J.\ Nucl.\ Phys.\ {\bf 28} (1978) 822 [Yad. Fiz. {\bf 28} (1978) 1597].










\bibitem{Drukker:2012de}
  N.~Drukker,
  JHEP {\bf 1310} (2013) 135
  [arXiv:1203.1617 [hep-th]].



\bibitem{Correa:2012hh}
  D.~Correa, J.~Maldacena and A.~Sever,
  JHEP {\bf 1208} (2012) 134
  [arXiv:1203.1913 [hep-th]].


\bibitem{Bajnok:2012xc}
  Z.~Bajnok, R.~I.~Nepomechie, L.~Palla and R.~Suzuki,
  ``Y-system for Y=0 brane in planar AdS/CFT,''
  JHEP {\bf 1208} (2012) 149
  [arXiv:1205.2060 [hep-th]].


\bibitem{Hegedus:2016eop}
  A.~Hegedus and J.~Konczer,
  ``Strong coupling results from the numerical solution of the quantum spectral curve,''
  arXiv:1604.02346 [hep-th].

\end{thebibliography}
\end{document}